\documentclass[12pt]{article}
\usepackage{amstex}
\begin{document}

\begin{center}
{\Large Membrane's Discrete Spectrum and BRST Residual 
Symmetry in the Conformal Light Cone Gauge.}
\vskip .5cm
{\large R. Torrealba.}
\vskip.5cm
{\it Depto. de Matematicas , Decanato de Ciencias,
Universidad Centro Occidental  "Lisandro Alvarado"
Barquisimeto  400, Venezuela.}

\end{center}

\begin{center}
{\bf Abstract}
\end{center}

\begin{quotation}

{\it In this work we present a proof of the discretness of the 
spectrum for bosonic 
membrane, in a flat minkowski space. This may be useful to show the 
quantum 
mechanical consistence 
of others bosonics extended models. This proof includes the BRST 
residual symmetry and was directly 
performed over the discretized membrane model. The BRST residual 
invariant 
effective action is explicity contructed.}

\end{quotation}

\section{Introduction}

Recently there has been a renewed interest in the theories of 
Super-p-branes, 
specially due to the duality relations between some Super-p-branes, 
Dirichlet Branes and 
Superstring in several dimension \cite{polchi}. These duality 
relation 
are usually stablished at different coupling limits 
(strong-weak) and
different compactification limits (large-short radios) 
of low energy phenomenological actions \cite{school}. There are also 
remarkable
results for the BPS spectra, and calculations of entropy for 
these new membrane theories. \cite{BPS} 

Due to the new membrane theories,
the problem of the supermembrane spectrum comes out as a matter 
of intense research . Some years ago
it was probed that the spectrum of the supermembrane was continuous 
whitout gap
from zero in flat space time  \cite{SUn}. More recently it has been 
claim \cite{Russo} that by compactifying on a torus some of the 
spatial dimensions the 
spectrum renders to be discrete, but others authors \cite{dewit} 
disagrees and obtains
still obtain a continuous spectrum. In a recent paper \cite{us} it is shown 
that both anwsers are 
correct depending on the irreducibility number of the wrapping on the torus.
	
In contrast with the extense work devoted to the Supermembranes,
 there exist relatively few works about the quantization of the 
bosonic membrane \cite{henn} \cite{odin}
 most of them are semiclassical aproximations and indicate a discrete 
spectrum 
. It is widely acepted that the membrane spectrum is 
discrete, but 
the in  \cite{kaku}\cite{Russo} it has pointed out that it 
seems to be continuous, 
although in these cases, the authors claim that for physical reasons 
(that involves the confinement
of wave fronts due to the uncertainty principle) the spectrum turns 
out to be discrete. So 
the objetive of this paper is to present a proof of 
the discretness of the bosonic membrane spectrum, from a 
BRST functional integral point of view.

The organization of this paper is the following: in section 2 we 
review the 
classical membrane theory, its invariances, constraints and residual 
gauge 
symmetry in the confomal light cone gauge fixing. In section 3 we 
explain the 
discretized membrane model and explicity construct the invariant 
residual 
BRST effective action. In section 4 we present a 
proof of the discretness of the 
bosonic membrane spectrum, taking into account the residual BRST 
gauge 
invariance. 

\section{Gauge Fixing and Residual Gauge Group}

We will start from the usual second order action 

\begin{equation}
S=-\frac{1}{8\pi^2\beta}\int_B\sqrt{-g}(g^{ab}x^{\mu},_ax_{\mu},_b-1)d^3\sigma\nonumber
\label{eq:1}
\end{equation}

\noindent{}which is equivalent to the Nambu-Goto-Dirac action 
\cite{accion} over the field equations.

The classical hamiltonian and constraints can be obtained
from (\ref{eq:1}), following Dirac's procedure and using
the ADM parametrization:
 
\begin{equation}
{\cal H}=\frac{N}{2\sqrt{\gamma}}(p^2+\gamma)+N^i(px,_i)
\label{eq:2}
\end{equation}

\noindent{}where $\frac{N}{2\sqrt{\gamma}}$ and $N^i$ are the 
Lagrange 
multipliers associated with the 3-d diffeomorphism generating
first class constraints: 

\begin{eqnarray}
&\phi=\frac{1}{2}(p^2+\gamma)&\nonumber\\
&\phi_i=px,_i\
\label{eq:3}
\end{eqnarray}

\noindent{}The conformal gauge fixing for this theory is defined as 
in \cite{fuji}

\begin{eqnarray}
&g_{oj}= 0&\nonumber\\
&g_{oo} + \gamma = 0&\
\label{eq:4}
\end{eqnarray}
 
It was shown some time ago \cite{hope} that even though this
gauge fixes the Lagrange multipliers, there is still a residual gauge 
group that allow us to fix the light cone gauge (LCG). This residual 
gauge group has as parameters the solutions to the homogeneous 
equations 
that arise from the gauge invariance of the gauge fixing conditions, 
i.e. 
the solutions to: 
  
\begin{eqnarray}
\delta{}(g_{oo}+\gamma)=2[-\partial_o\zeta_{res}^o+\partial_i\zeta^i_{res}] =0\\
\label{eq:5}
\delta{}(g_{oi})=\gamma_{ij}\partial_o\zeta_{res}^i+\gamma
\partial_i\zeta^o_{res} = 0
\label{eq:6}
\end{eqnarray}

The temporal evolution of $x^+$ is given by the field equation:
$\partial_{oo}x^+=0$, that could be directly integrated as:
$x^+=c^+\tau+\kappa(\sigma^1,\sigma^2)$ where the function
$\kappa(\sigma^1,\sigma^2)$ could be determined using the residual 
gauge 
parameters
$\zeta^i_{res}$. Note that $\delta{}x^+=c^+\tau$ fixes the residual 
gauge 
parameter $\zeta^o_{res}$, so:

\begin{eqnarray*}
	\partial_o \zeta^i_{res} =  0 &&  \\
	 & & \Rightarrow \:\zeta^i_{res} = \epsilon^{ij} \partial_j 
f(\sigma^1,\sigma^2) \\
\partial_i \zeta^i_{res} =  0 && 
\end{eqnarray*}

The LCG fixing allows us to determine the $d-2$ transverse part, 
explicitly solving the constraints we get:

\begin{eqnarray}
\partial_ix^-=\frac{1}{c^+}\partial_i\vec{x}.\vec{p}\label{eq:7}\\
p^+=c^+\label{eq:8}\\
p^-=\partial_0x^-\label{eq:9}
\end{eqnarray}

These equations do not
exhaust the content of the constraints (\ref{eq:3}). Indeed, if we 
take 
the {\it 2d curl} of $\phi_i$, we get the residual constraint 
\cite{sezgin}

\begin{equation}
T=\epsilon^{ij}\vec{p},_i\vec{x},_j=
\vec{p},_1.\vec{x},_2-\vec{p},_2.\vec{x},_1
\label{eq:10}
\end{equation}

that generates the residual group. This group is the subgroup of 2 
dimensional difeomorphisms that preserve areas, and its generator has 
a 
closed first class algebra, namely:
 
\begin{equation}
\{T(\sigma),T(\bar{\sigma})\}=\epsilon^{ij}
\partial_jT(\sigma)\partial_j\delta(\sigma-\bar{\sigma})
\label{eq:11}
\end{equation}

The action of this group on the canonical variables is given as 
follows:
\begin{equation}
\delta_Tx=\{x(\sigma),\int{}d^2\bar{\sigma}\lambda\bar{\sigma}
T(\bar{\sigma})=\xi^j\partial_jx(\sigma)
\label{eq:12}
\end{equation}
\begin{equation}
\delta_Tp=\{p(\sigma),\int{}d^2\bar{\sigma}\lambda\bar{\sigma}
T(\bar{\sigma})=\partial_j(\xi^jp(\sigma))
\label{eq:13}
\end{equation}

where $\xi^i\equiv\epsilon^{ij}\partial_j\lambda$ may be identified 
with
$\zeta^i_{res}$. According to the above formulas, the coordinates $x$ 
transform
as a scalars while their corresponding momenta $p$ as scalar densities
in two dimensions as expected.

Performing variations of the action (\ref {eq:1}) with respect to the 
variables
 $g_{ab}$, and using (\ref {eq:4}) we get
 
\begin{eqnarray}
&\partial_ax^-=-\partial_0x^I\partial_ax^I&
\label{eq:14}\\
&\partial_0x^-=-\frac{1}{2}(\partial_0x^I)^2-
\frac{1}{2}det(\partial_ax^I\partial_bx^I)&
\label{eq:15}
\end{eqnarray}

where  we denoted by $x^I$  the Light cone transverse part of $x^m$. 
These 
equations allow us to solve the minus  sector $x^-$ .

The equations for the  d-2 transverse sector, may be obtained the 
following effective transverse action

\begin{equation}
L=\frac{1}{2}(\partial_0x^I)^2-\frac{1}{4}det(\partial_ax^I\partial_bx^I)
\label{eq:16}
\end{equation}

this could be rewritten as

\begin{equation}
L=\frac{1}{2}(\partial_0x^I)^2-\frac{1}{4}
\{x^I,x^J\}_{\cal LB}\{x^I,x^J\}_{\cal LB}
\label{eq:17}
\end{equation}

where  the Lie bracket is  definided by

\begin{equation}
\{A,B\}_{\cal LB}\equiv\epsilon^{ij}\partial_iA\partial_jB
\label{eq:18}
\end{equation}

	The action (13) is invariant under gauge transformations generated 
by the 
constraints (3) but with the parameters $\xi^i= 
\epsilon^{ij}\partial_j \lambda$,
 the transformation of $ x $ by the Lie brackets follows from

\begin{equation}
\delta{}x(\sigma)=\{x(\sigma),\lambda(\sigma)\}_{\cal LB}
\label{eq:19}
\end{equation}

that is equivalent to the residual gauge invariance generated by  
$T$  
through

\begin{eqnarray}
\delta_Tx&=&\{x(\sigma),\int{}d^2\bar{\sigma}\lambda(\bar{\sigma})
T(\bar{\sigma})\}_{\cal PB}\nonumber\\
\\
&=&\{x(\sigma),\lambda(\sigma)\}_{\cal 
LB}=\zeta^j\partial_jx(\sigma)\nonumber
\label{eq:20}
\end{eqnarray}

\begin{equation}
\mbox{where: }\zeta^j\equiv\epsilon^{ij}\partial_j\lambda(\sigma)
\label{eq:21}
\end{equation}

	The  transverse action (\ref {eq:17}) has just the structure of the 
action 
for a  Yang Mills theory compactified to one dimension, in the 
Coulomb gauge \cite{SUn}. 
This equivalence is a particular characteristic of the 2-brane and 
could not be easilly 
extended to other p-branes. Altough we still have to discuss the 
residual gauge 
symmetry of the membrane generated by the residual constraint 
(\ref{eq:10}) that in principle is 
absent from a Yang Mills theory.

\section{Discretized membrane model.}

Introducing a base of functions over the section of B at constant 
time \cite{SUn}

\begin{eqnarray}
	x^I(\tau,\sigma^i) & = & x^I_o (\tau,\sigma^i) + 
\sum_{A}^{}x^{IA}(\tau)  
Y_A (\sigma^1,\sigma^2),
	\label{eq:22}\\
	p^I(\tau,\sigma^i) & = & p^I_o (\tau,\sigma^i) + 
\sum_{A}^{}p^{IA}(\tau)  
Y_A (\sigma^1,\sigma^2)
	\label{eq:23}
\end{eqnarray}

the Hamiltonian is obtained from (17) using  $D_ox^I=\partial_o x^I = 
p^I$ 

\begin{eqnarray}
& H = \frac{1}{2c^+} [p^I_o p^I_o + p^{IA} p^{IA}] 
+\frac{1}{4}[f_{ABC} x^{IA} x^{JB}]^2 &
\label{eq:24}\\
	& f_{ABC} = \int d^2 \sigma  Y_A \{ Y_B,Y_C\} &
	\label{eq:25}
\end{eqnarray}
where $f_{ABC}$ are Lie algebra structure functions analogs.

	To obtain a correct theory of discretized membranes we must impose 
the
residual constraint (\ref{eq:10}) of the membrane, this implies a set 
of 
constraints over our discretized membrane model that are the 2-brane 
analogous of Virasoro 
constraints.

\begin{eqnarray*}
	&T =\{ P_I,x^I\} =(x^{IA}p_I^B) f_{AB}^C Y_C =0& \\
	&\Rightarrow L_A = f_{ABC} x^{IB} p_I^C  = 0&
\end{eqnarray*}

	We may define a first quantization theory for the discretized 
membrane, 
where the Hilbert space consists of the scalar wave functions 
valuated 
over the infinite set of coeficients  $x^{IA}(t) (A=1,...,\infty${ 
}$I= 1,...,d-2 )$ 
instead of  $x^{IA}(\tau, \sigma^1, \sigma^2)$.

\begin{equation}
	\Phi(x^{IA}): \Bbb{R}^{N(d-2)} \rightarrow  \Bbb{C}
	\label{eq:26}
\end{equation}

	The operators position and momentum are defined in the Schr\"{o}dinger 
representation as

\begin{equation}
	X^{IA} |\Phi>  = x^{IA} |\Phi> \quad \mbox {and} \quad P_{IA} |\Phi> 
= 
	-i\frac{\partial}{\partial x^{IA}} |\Phi>
	\label{eq:27}
\end{equation}

	Eliminating the zero mode from (\ref{eq:24}) we get the Schr\"{o}dinger 
equation 
	
	\begin{equation}
		\left[-\frac{1}{2}(\frac{{\partial}^{2}}{{\partial x^{IA}_{0}}^{2}} 
+(\frac{1}{2}f_{ABC} 
		x^{IB} x^{JC})^2 \right]|\Phi> = E |\Phi>
		\label{eq:28}
	\end{equation}

that jointly with the residual constraints $L_{A}$ are the equations for 
the 
wave functions.

\begin{equation}
	-if_{ABC} x^{IA} (\frac{\partial}{\partial x^{IA}}) |\Phi> = 0
\label{eq:29}
\end{equation}

	From these equations it is evident that the theory is not uniquely 
defined. For example consider a membrane with periodic boundary 
conditions 
$x(\sigma^1) = x(\sigma^1 + 2\pi k/m)$  and $x(\sigma^2) = x(\sigma^2 
+ 2\pi k/n)$.
 
 A complete set of functions is

\begin{equation}
	Y_{mn} = exp(im\sigma_1 + in\sigma_2)
	\label{eq:30}
\end{equation}

and the structure constants are 

\begin{equation}
	f_{ABC} = f_{mn,pq,rs} = (A\times B) \delta ^{A+B}_C
	\label{eq:31}
\end{equation}

where  A=(m,n);  B=(p,q)  and  C=(r,s)  $\in$	$Z^2$ that coincides 
with 
the structure 
constants of a NxN matrix realization of $SU(N)$ in the  $N 
\rightarrow \infty$ 
limit \cite{SUn}.

	It is easy to see that  coeficients  $L_A$ of the constraint $T$ 
satisfy 
the same Poisson algebra, that the base (\ref{eq:30}) in term of Lie 
Bracket. In fact

\begin{eqnarray*}
	&\{L_A,L_D\}_P = [f_{ABC}f_{DCF}- f_{DBC}f_{DCF}] x^B p^F &  
\end{eqnarray*}

but $T_A = f_{A(BC)}$  correspond to the adjunt representation , that 
satisfy  

\begin{equation}
	[f_{A(BC)},f_{D(CF)}] = f_{ADE}f_{E(BF)} \quad \mbox{then} \quad 
\{L_A,L_D\}_P = f_{ADF}L_F
	\label{eq:32}
\end{equation}

The BRST generating charge for a closed constraint algebra 
\cite{brst} is 

\begin{equation}
	\Omega = c^AL_A -\frac{1}{2}c^Ac^Bf_{AB}^C\mu _C
	\label{eq:33}
\end{equation}

Following a modified BFV \cite{bfv} approach The  Functional Integral 
incluiding 
the BRST invariant terms is given as:

\begin{eqnarray}
	&{\large I} =  \int Dz \: e^{\int dt\: p\dot{x} -\mu_A \dot{c}^A - 
H_{o(brst)} - \delta (\lambda^A\mu_A) +\delta(\underline{c}_A 
\chi^A)} & \
	\label{eq:34} \\
	 & Dz=Dp\,Dx\,D\mu\,Dc\,D\underline{c}\,DB\,D\Theta& \nonumber\
	\label{}
\end{eqnarray}

where $p, x$ and $\mu, c$ are canonical variables, while the others 
variables are not canonical.

	The BRST invariant Hamiltonian is given by
	
\begin{equation}
	{\cal H}_{o(brst)} = {\cal H}_{o} + \mu_a\,{}^{(1)}V_B^A \,c^B
	\label{eq:35}
\end{equation}

where $V_B{}^A$ are the coeficients of the commutator 

\begin{equation}
	\{L_B(\sigma),{\cal H}_{o}(\underline{\sigma})\}= {}^{(1)}V_B^A \, 
L_A
	\label{eq:36}
\end{equation}

we obtain that this coeficients are null in virtue of

\begin{equation}
	\{ Y^FL_F, {\cal H}_{o}\} =\{ T,{\cal H}_{o}\} = 
	\epsilon^{ij}\partial_i\{ \phi_j(\sigma),\phi_3(\underline{\sigma)} 
\} = 0
	\label{eq:37}
\end{equation}

this implies that  $V_B{}^{A} = 0 {  } \forall {   } A $ and ${ }B $. 
So we get	

\begin{equation}
	{\cal H}_{o(brst)} = {\cal H}_{o} = \frac {1}{2}\,p^{IA}.p^{IA} + 
\frac{1}{4}\,f_{AB}^C\,f_{CED}\,x^{IA}x^{JB}x^{IE}x^{JD}
	\label{eq:38}
\end{equation}

The transformation laws of a object depending on canonical variables 
are 
given by 

\begin{equation}
	\delta F(p,x,\mu,c) = \{ F,\Omega\}
	\label{eq:39}
\end{equation}

while the non canonical variables the transformation laws are given by

\begin{eqnarray}
	\delta \underline{c}_a = B_A & \qquad & \delta B_A = 0
	\nonumber\\
	\delta \lambda^A = \Theta ^A & \qquad  & \delta \Theta ^A = 0
	\label{eq:40}
\end{eqnarray}

	Using the transformation laws and the Hamitonian (\ref{eq:38}) into 
(\ref{eq:34}) we get

\begin{eqnarray}
	&{\large I} =  \int Dz \: e^{\int dt\: p\dot{x} -\mu_A \dot{c}^A - 
H_{o(brst)} - 
	\lambda^A (L_A - c^B f_{AB}^C\mu_C) + B_A\chi^A -\underline{c}_A 
\delta 
	\chi^A} & \
	\label{eq:41} \\
	 & Dz=Dp\,Dx\,D\mu\,Dc\,D\underline{c}\,D\lambda \,DB\,D\Theta& 
\nonumber\
	\label{}
\end{eqnarray}

note that $H_o{}^{(brst)}$ is the BRST invariant Hamiltonian and not 
the 
BRST effective Hamiltonian, the former which be deduced from the 
effective 
action in the functional integral after fixing the gauge.

	First we will integrate in the $\lambda$ variables and get 
functional 
deltas over the BRST extended constraints

\begin{equation}
	L_{A(brst)} = L_A - c^B f_{AB}^C\,\mu_C
	\label{eq:42}
\end{equation}

that generates the same constraint algebra (\ref{eq:32}) than $L_A$	

\begin{equation}
	\{ L_{A(brst)}, L_{D(brst)}\} = f_{ADF} \, L_{F(brst)}
	\label{eq:43}
\end{equation}

	We will now fix the residual gauge freedom generated by the 
constraints 
taking

\begin{equation}
	\chi^C = \lambda^C - \kappa ^C
	\label{eq:44}
\end{equation}

where $\kappa^C$ is a suitable collection of constants so  $\delta 
\kappa^C = \theta_C$ .
Integrating (\ref{eq:41}) in the auxiliary variables $B$ and $x^{IA}$ we 
obtain

\begin{eqnarray}
	&{\large I} =  \int Dz \: \delta(L_{A(brst)})  \delta(\lambda^C - 
\kappa^C) \delta(\underline{c}_A - \mu_A) e^{\int dt\: p\dot{x}-\mu_A 
\dot{c}^A - H_{o(brst)}} & \
\label{eq:45} \\
	 & Dz=Dp\,Dx\,D\mu\,Dc\,& \nonumber\
	\label{}
\end{eqnarray}

the last term in the exponential is the effective Hamiltonian

\begin{equation}
	{\cal H}_{eff} = {\cal H}_{o(brst)} = \frac {1}{2}\,p^{IA}.p^{IA} + 
\frac{1}{4}\,f_{AB}^C\,f_{CED}x^{IA}x^{JB}x^{IE}x^{JD}
	\label{eq:46}
\end{equation}

that only in this gauge choice coincides with the BRST invariant 
Hamiltonian, 
but submited to the restrictions implied by the deltas in the 
functional 
integral, they are:	

\begin{equation}
	\underline{c}_A = \mu_A \quad \lambda^A = \kappa ^A \quad \mbox{and} 
\quad L_A - c^B f_{AB}^C \,\mu_C = 0
	\label{eq:47}
\end{equation}

\section{Discretre spectrum of the membrane.}

	 In this section we will probe that the spectrum of the membranes is 
discrete, 
performed directly on de discretized membrane model. We also take 
into 
account 
the local constraints $L_A$ and the residual BRST invariance they 
generate 
in a fix 
gauge choice, but the BFV theorem \cite{brst} garantized that our 
result is gauge 
independient and our proof have the advantage that we never take the 
limit 
$N \rightarrow \infty$ 
of the group $SU(N)$.

	We will use a corollary due to B. Simon \cite{simon} of a beautiful 
theorem due to  
Fefferman and Phong \cite{fefferman} about the spectral dimension of 
the quantum 
Hamiltonian. This Corollary stablishes that the number of eigenvalues 
(counting 
multiplicities) of the Hamiltonian is finite for every finite total 
system 
energy value, if the Hamiltonian  operator for the quantum system is

\begin{equation}
	{\cal H} = -\nabla^2 + V(X), \quad x\in {\cal R}^M \quad \mbox{and} 
\quad 
V(x)\geq 0
	\label{eq:48}
\end{equation}

and the potential $V(x)$ can be written as a sum of homogeneous 
polynomials of degree 2

\begin{equation}
	V(x) = \sum_{j=1,..,m}Q_J^2 , \quad \mbox{that satisfies} \quad 
	\sum_{j=1,..,m\,;\, \alpha = a,..,n} \left(\frac{\partial 
Q_j}{\partial 
x^{\alpha}}\right)^2 \geq 0
	\label{eq:49}
\end{equation}

So we are going to probe that the effective BRST Hamiltonian 
accomplishes 
all these conditions.

	From (\ref{eq:46}) it is evident that the potential could be written 
as

\begin{equation}
	V(x^{IB}) = \sum_{A,i,j} (Q^A_{ij})^2
	\label{eq:50}
\end{equation}

where $Q^C_{ji} = f_{ABC} x_i^A x_j^B$ are homogeneous polinomials of 
degree 2. It is also evident that  $V(x) \geq 0$. 

The left hand side of (\ref{eq:50}) is given by

\begin{equation}
	\sum_{A,I,E} x^{IA}\,f_{ABC}\,f_{EBC}\,x_I^E = (x^I,x_I)_K
	\label{eq:51}
\end{equation}

\noindent
as $T_A = f_{A(BC)}$ correspond to the adjoint representation of 
(\ref{eq:32}) 
then  $x^I = x^{IA} f_{A(BC)}$ 
are forms valued on the adjoint representation and the product in 
(\ref{eq:51}) 
correspond to the usual definition of the Killing product.

	We only have to probe that this Killing product is not negative when 
$x^I \neq 0$

\begin{equation}
	(x^I,x_I)_K = x^{IA} x_I^E K_{AE}
	\label{eq:52}
\end{equation}

where the Killing metric is diagonal 	 $K_{AE} = (A x B)(E x B) 
\delta_{A+B,E+B}$ 
which implies that 	$K =  tr (T_AT_E) =  \sum_{A}\sum_{B} (A x B)^2  >  0 $  
then the 
Killing product (\ref{eq:52}) is positive definite when $x^I \neq 0$

\begin{equation}
	(x^I,x_I)_K = K.\eta_{IJ}x^{IA}x^{JA}> 0
	\label{eq:53}
\end{equation}

\noindent because the light cone metric $\eta^{IJ}$ is positive and 
$(x^I , x^I)_K $ 
is a sum of positive terms. 

So this  discrete membrane model accomplishes all the above 
conditions then we conclude that the spectrum is discrete for every 
finite amount of energy.

\section{Conclusions.}

	In this paper we obtain the BRST effective Hamiltonian for the 
membrane 
in a gauge fixing, namely the conformal light cone gauge plus the 
residual 
gauge fixing conditions. We conclude that the effective Hamiltonian 
satisfies the conditions of the Simon and Fefferman and Phong 
theorems, this means that for 
finite energy the spectrum of the membrane is discrete and finite. 
Althought this 
result was obtained in a particular gauge fixing, due to the BFV 
theorem the 
result must be valid in all gauges.

	In Physical terms, this is a example of a potential that is non 
confinant 
for a system of classical particles but that is quantically confinant 
for wave 
functions due to the uncertanty principle. 

	The way in which Supersymmetry breaks this result must be study 
carefully to include the residual Symmetry, allowed gauge fixing 
conditions, and global constraints.

\end{document}